\documentclass[twocolumn,apl,amsfonts,amsmath,amssymb,floatfix,superscriptaddress]{revtex4-1} 
\usepackage{color}
\usepackage{soul}
\usepackage{hhline}	
\usepackage{mathrsfs}
\usepackage{graphicx}
\usepackage{dcolumn}
\usepackage{bm}
\usepackage{multirow}
\usepackage{booktabs}
\usepackage{afterpage}
\usepackage{amsmath}
\usepackage{braket}
\usepackage{array}

\begin{document}

\title{Thermoelectric properties of (Ba,K)Cd$_{2}$As$_{2}$ 
crystallized in the CaAl$_2$Si$_2$-type structure}

\author{H. Kunioka}
\affiliation{National Institute of Advanced Industrial Science and Technology (AIST), Tsukuba, Ibaraki 305-8568, Japan.}
\author{K. Kihou}
\affiliation{National Institute of Advanced Industrial Science and Technology (AIST), Tsukuba, Ibaraki 305-8568, Japan.}
\author{H. Nishiate}
\affiliation{National Institute of Advanced Industrial Science and Technology (AIST), Tsukuba, Ibaraki 305-8568, Japan.}
\author{A. Yamamoto}
\affiliation{National Institute of Advanced Industrial Science and Technology (AIST), Tsukuba, Ibaraki 305-8568, Japan.}
\author{H. Usui}
\affiliation{Department of Physics, Osaka University, Toyonaka, Osaka 560-0043, Japan}
\author{K. Kuroki}
\affiliation{Department of Physics, Osaka University, Toyonaka, Osaka 560-0043, Japan}
\author{C. H. Lee}
\email{c.lee@aist.go.jp}
\affiliation{National Institute of Advanced Industrial Science and Technology (AIST), Tsukuba, Ibaraki 305-8568, Japan.}
\date{\today}
\begin{abstract}
As-based Zintl compounds Ba$_{1-x}$K$_x$Cd$_2$As$_2$ 
crystallized in the CaAl$_2$Si$_2$-type structure (space group $P\bar{3}m1$) 
were prepared using solid-state reactions followed by hot-pressing. 
We have successfully substituted K for Ba up to $x$ = 0.08, producing hole-carrier doping with concentrations up to 1.60$\times$10$^{20}$ cm$^{-3}$. 
We have determined the band-gap value of non-doped BaCd$_2$As$_2$ to be 0.40 eV from the temperature dependence of the electrical resistivity. 
Both the electrical resistivity and the Seebeck coefficient decrease with hole doping, leading to a power factor value of 1.28 mW/mK$^2$ at 762 K for $x$ = 0.04. 
A first-principles band calculation shows that the relatively large power factor mainly originates from the two-fold degeneracy of the bands 
comprising As $p_{x,y}$ orbitals and from the anisotropic band structure at the valence-band maximum. 
The lattice thermal conductivity is suppressed by the K doping to 0.46 W/mK at 773 K for $x$ = 0.08, presumably due to randomness. 
The effect of randomness is compensated by an increase in the electronic thermal conductivity, 
which keeps the total thermal conductivity approximately constant. 
In consequence, the dimensionless figure-of-merit $ZT$ reaches a maximum value of 0.81 at 762 K for $x$ = 0.04.
\end{abstract}

\maketitle

\section{Introduction}
High-performance thermoelectric materials are desirable for energy conservation, 
as they can generate electrical power from waste heat. 
Intense efforts have been devoted to improving the performance of such materials, 
but it is still not high enough for practical applications. 
Thermoelectric performance is characterized by the dimensionless figure-of-merit 
$ZT = S^2T/\rho\kappa$, where $S$ is the Seebeck coefficient, $\rho$ is the electrical resistivity, 
and $\kappa$ is the total thermal conductivity. 
The difficulty in improving material performance comes from the conflicting requirements to 
exhibit high electrical conductivity while keeping the thermal conductivity as low as possible. 

\begin{figure}
\includegraphics[width=\columnwidth]{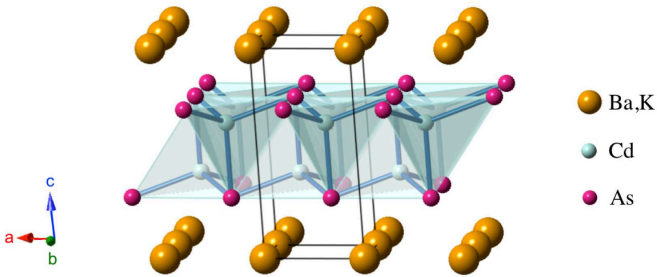}
\caption{\label{crystal} Crystal structure of (Ba,K)Cd$_{2}$As$_{2}$ with the space group $P\bar{3}m1$.}
\end{figure}

Many new thermoelectric materials have been discovered in recent decades. 
Among them, the 122 Zintl compounds are a promising system. \cite{Kauzlarich_rev,Guo_rev,Shuai_rev,Gascoin_(CaYb)Zn2Sb2,May_(CaEuYb)Zn2Sb2,Shuai_(CaEuYb)Zn2Sb2,Wood_Ca(ZnMg)2Sb2,Zhang_Eu(ZnCd)2Sb2,Zhang_EuCd2Sb2,Zhang_(YbEu)Cd2Sb2,Cao_(CaYb)Cd2Sb2,Yu_Yb(ZnMn)2Sb2,Guo_Yb(CdMn)2Sb2,Song_Mg(MgAg)2Sb2,Aydemir_Ba(GaZn)2Sb2,Aydemir_(BaNaK)Ga2Sb2,Tamaki_Mg3Sb2,Zhang_Mg3Sb2,Wang_Yb(ZnCd)2Sb2,Toberer_AZn2Sb2,Shuai_(EuCaYb)Mg2Bi2,Shuai_(CaYb)Mg2Bi2,May_AMg2Bi2}
In particular, Mg$_3$Sb$_2$ exhibits a $ZT$ value of 1.65 at 725 K, with n-type electronic conductivity, 
where the Mg atoms are located at two atomic sites in the ratio of 1 : 2 so that it is categorized as 122 system. \cite{Tamaki_Mg3Sb2,Zhang_Mg3Sb2} 
As for the p-type materials, Yb(Zn,Cd)$_2$Sb$_2$ and (Ca,Yb,Eu)Mg$_2$Bi$_2$ 
exhibit $ZT$ values of 1.2 at 700 K and 1.3 at 873 K, respectively. \cite{Wang_Yb(ZnCd)2Sb2,Toberer_AZn2Sb2,Shuai_(EuCaYb)Mg2Bi2,Shuai_(CaYb)Mg2Bi2,May_AMg2Bi2}
Those compounds crystallize in the CaAl$_2$Si$_2$-type structure \cite{Brechtel_CaAl2Si2,Klufers_CaAl2Si2} with the space group $P\bar{3}m1$ (Fig. \ref{crystal}). 
Characteristically, they consist of alternately stacked anionic layers and cation sheets. 
The anionic layers form two-dimensional networks of edge-sharing tetrahedrons with covalent bonding. 

The 122 Zintl compounds typically exhibit quite low lattice thermal conductivities ($\kappa\rm_L$) less than 1 W/mK 
at high temperatures, although their crystal structure is relatively simple. 
The origin of the low $\kappa\rm_L$ is still controversial. 
One hypothesis attributes it to the existence of lone pairs around the pnictogen atoms. \cite{Toberer_AZn2Sb2}
Actually, lone pairs are present in various thermoelectric materials. 
For example, it has been proposed that lone pairs are responsible for the extremely low $\kappa\rm_L$ in SnSe. \cite{SnSe_C.S.Li}
On the other hand, the high power factors ($S^2/\rho$) of the 122 Zintl compounds can originate from multi-valley band structures in n-type, \cite{Tamaki_Mg3Sb2,Zhang_Mg3Sb2}
or from orbital degeneracy at valence-band edges in p-type. \cite{Zhang_122calc}
The wide variety of 122 Zintl compounds offer the advantage of allowing the design of ideal band structures. 


\begin{figure}
\includegraphics[width=\columnwidth]{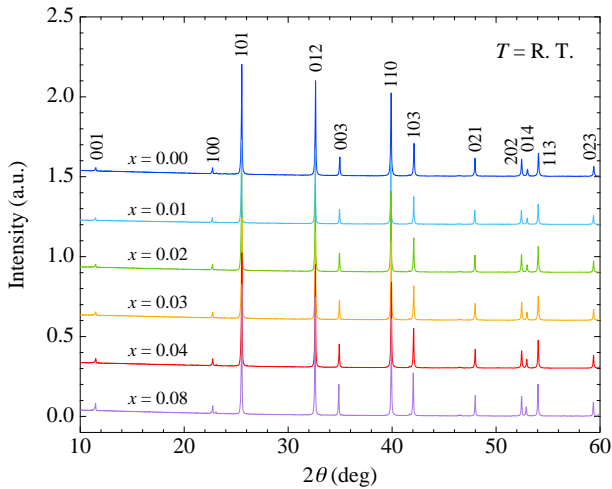}
\caption{\label{XRD} X-ray diffraction patterns for Ba$_{1-x}$K$_x$Cd$_2$As$_2$ ($x \leq$ 0.08) at room temperature. 
Typical indices of the reflections are described.}
\end{figure}

To date, the 122 Zintl compounds have been intensively explored for Sb-based compounds. 
In contrast, compounds with lighter pnictogen atoms such as As and P have been less studied, 
because thermal conductivity generally increases with lighter atoms. \cite{Ponnambalam_Ca(ZnMnCu)2P2,Ponnambalam_YbCuZnP}
Recently, we have found that (Ba,K)Zn$_2$As$_2$ exhibits quite a low value of $\kappa\rm_L$ = 0.8 W/mK at 773 K. 
This is comparable with Sb-based compounds, leading to a $ZT$ value of 0.67. \cite{Kihou_BKZA} 
Although the crystal structure of (Ba,K)Zn$_2$As$_2$ is the $\alpha$-BaCu$_2$S$_2$-type \cite{Klufers_BaCu2S2}
different from the CaAl$_2$Si$_2$-type, 
this suggests that As-based compounds can also be promising thermoelectric materials. 
However, there have been few studies of As-based 122 Zintl compounds. 
In particular, there are no reports of As-based compounds with the CaAl$_2$Si$_2$-type structure 
as high-performance thermoelectric materials, even though they are promising candidates. 
In this study, we have thus explored the thermoelectric properties of (Ba,K)Cd$_2$As$_2$ 
crystallized in the CaAl$_2$Si$_2$-type structure. 

\section{Experimental}
Polycrystalline samples of Ba$_{1-x}$K$_x$Cd$_2$As$_2$ were synthesized using solid-state reactions. 
First, BaAs, Cd$_3$As$_2$ and BaCd$_2$ were synthesized as precursors 
using Ba (3N, chunks), K (3N, chunks), As (6N, 1-5mm chunks), and Cd (6N, shot) as starting materials. 
They were mixed at stoichiometric compositions, loaded them into alumina crucibles to produce Cd$_3$As$_2$, BaCd$_2$, and KAs, 
and into a silica tube to create BaAs. 
The silica tube was heated at 750 $^\circ$C for 24 h in a box furnace. 
The alumina crucibles were encapsulated in a screw-top stainless-steel container,\cite{Kihou_KFA} heated at 800 $^\circ$C to produce Cd$_3$As$_2$ and BaCd$_2$ 
and at 650 $^\circ$C to produce KAs, maintaining those temperatures for 24 h. 
These precursors were then mixed with As at stoichiometric ratios, ground, and pressed into pellets at room temperature to synthesize Ba$_{1-x}$K$_x$Cd$_2$As$_2$. 
The pellets were placed in an alumina crucible and encapsulated in a screw-top stainless-steel container in an Ar atmosphere. 
Then, they were heated at 900 $^\circ$C for 24 h, followed by water quenching. 
All the sample treatments described above were conducted in a glovebox filled with dried Ar gas. 
To obtain dense pellets, hot pressing were performed. 
The pellets were ground, wrapped in expanded graphite sheets, and put into graphite dies. 
The graphite dies were heated at 850 $^\circ$C for 1 h while applying a uniaxial pressure of 70 MPa under an Ar gas flow. 
Finally, the dense pellets were annealed at 530 $^\circ$C for 3 h. 

Powder X-ray diffraction was conducted at room temperature using a SmartLab (RIGAKU) diffractometer. 
Cu K$\alpha_1$ radiation was used at 40 kV and 45 mA. 
The scattering angles (2$\theta$) ranged between 5 and 140 deg. 
and the scan speed for the 2$\theta$ was 0.2 deg. / min.
The obtained dense pellets were pulverized for these measurements. 
The diffraction patterns were analyzed by the Rietveld method using RIETAN-FP. \cite{Izumi2007}
The obtained parameters are given in the supplementary information. 
The densities of the hot-pressed samples were determined by the Archimedes method. 
The relative density of all of them was over 96 \%. 

\begin{figure}
\includegraphics[width=\columnwidth]{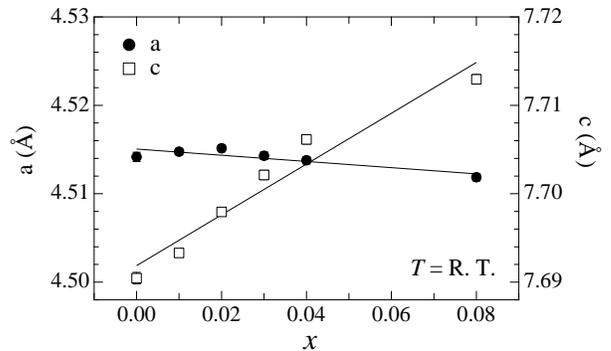}
\caption{\label{lattice const} Doping dependences of the lattice constants $a$ and $c$ evaluated through the Rietveld analysis.}
\end{figure}

The electrical resistivity and the Seebeck coefficient were measured using the four-probe method with a ZEM3 (ADVANCED-RIKO) instrument. 
Ag paste was used to secure contacts for the $x$ = 0.00 and 0.01 samples. 
The temperature ranged from room temperature to 762 K. 
Sample space was filled with 0.5 atm He gas. 
The dense pellets were cut into rectangular parallelepipeds with dimensions of approximately 2$\times$2$\times$8 mm$^3$. 
The measurements were conducted along the direction parallel to the uniaxial pressure axis which had been applied in the hot press. 

\begin{figure}
\includegraphics[width=\columnwidth]{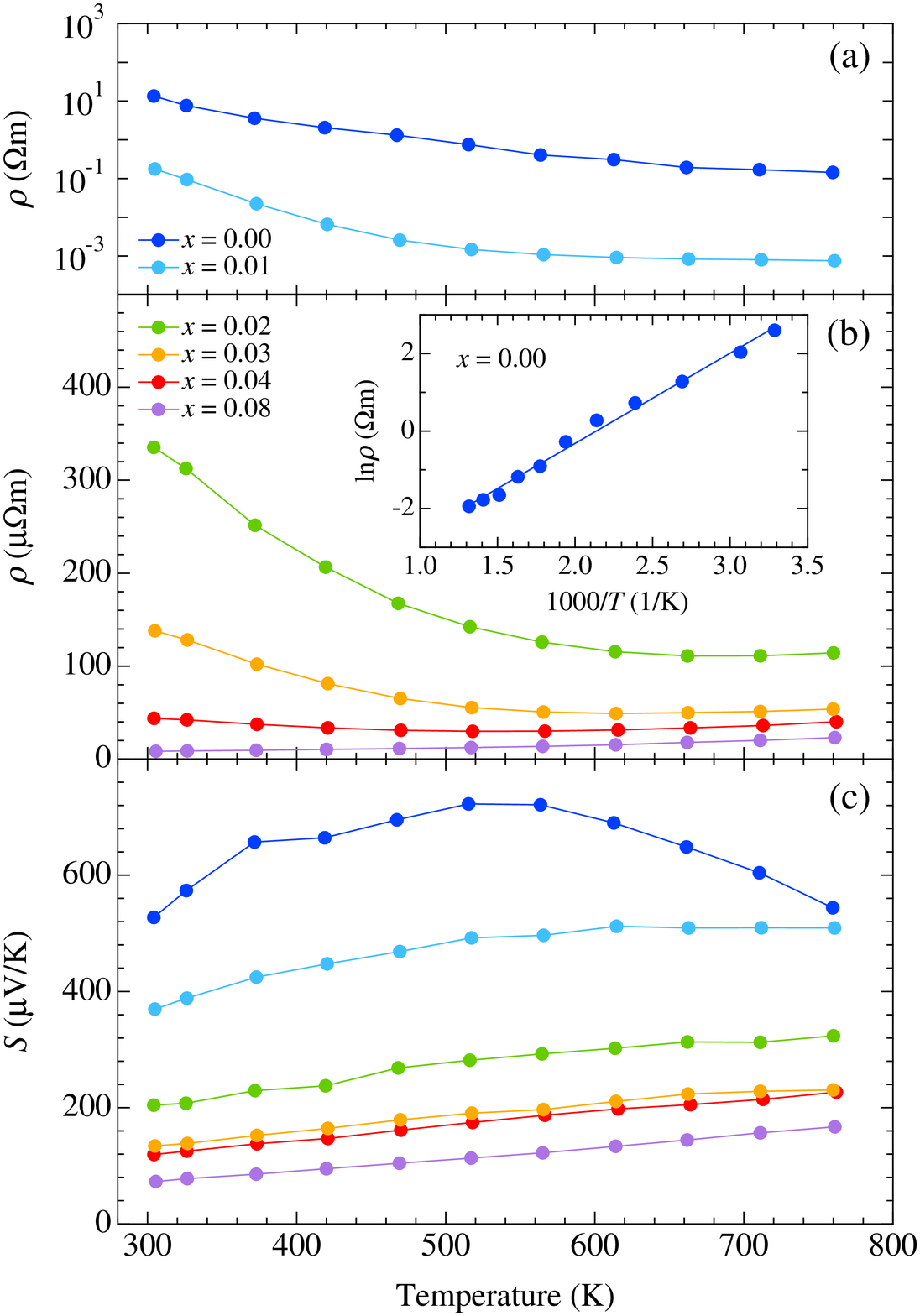}
\caption{\label{R_S} Temperature dependences of (a) resistivity for $x \leq$ 0.01, 
(b) resistivity for $x \geq$ 0.02, and (c) Seebeck coefficient. 
The inset in (b) shows an Arrhenius plot of the resistivity for $x$ = 0.00 with a linear fit.}
\end{figure}

\begin{figure}
\includegraphics[width=\columnwidth]{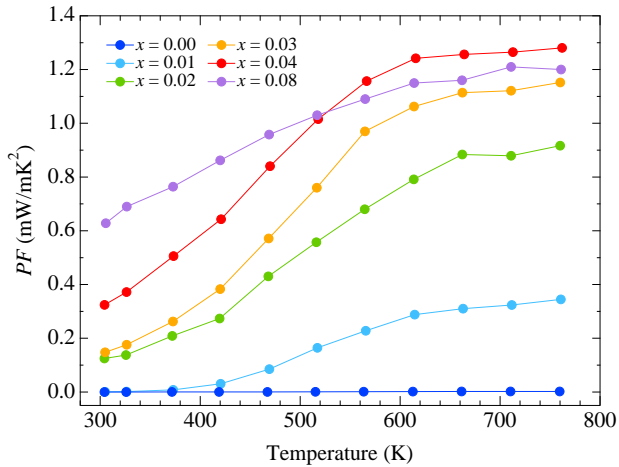}
\caption{\label{PF} Temperature dependence of the power factor for Ba$_{1-x}$K$_x$Cd$_2$As$_2$.}
\end{figure}

The thermal diffusivity ($D$) and the specific heat ($C\rm_p$) were measured by the laser-flash method using a LFA457 (Netzsch) instrument in an Ar gas flow. 
The size of used samples was 10 mm in diameter and 2 mm in thickness. 
The measurements were conducted along the direction parallel to the pressure axis applied in the hot press. 
The total thermal conductivity was calculated using the equation $\kappa$ = $DC\rm_p\it d\rm_s$, where $d\rm_s$ is the sample density. 
The specific heat of one of the samples was also measured by means of differential scanning calorimetry 
using a DSC404 F3 (Netzsch) instrument in an Ar gas flow to ensure the accuracy of the results measured with the laser-flash method. 

The Hall coefficient ($R\rm_H$) was measured at room temperature using the van der Pauw method in magnetic fields of 0.3 to 2 T. 
The shape of samples was a flat plate with typical dimensions of 6$\times$6$\times$0.7 mm$^3$. 
The magnetic field was applied in direction normal to the plane surface of the sample. 
The Hall carrier concentration ($n\rm_H$) was calculated from the equation $R\rm_H$ = 1 / $n\rm_H\it e$ where $e$ is the elementary charge.

A first-principles band-structure calculation for BaCd$_2$As$_2$ was performed using the WIEN2k package\cite{Blaha} with the modified Becke-Johnson exchange-correlation functional\cite{MBJ1,MBJ2} including spin orbit coupling.
The value of $RK_{\rm max}$ was set to 7, and we took 2,000 $k$-points for the self-consistent calculation and 10,000 $k$-points to calculate the density of states.
The Seebeck coefficient was obtained from Boltzmann transport theory implemented in the BoltzTraP code\cite{Madsen} with 100,000 $k$-points.
The relaxation time was taken to be an undetermined constant in the present study, and it cancels out in the calculation of the Seebeck coefficient.
The final result for the Seebeck coefficient was obtained by averaging over the $xx$, $yy$, and $zz$ components of the Seebeck coefficient tensor, which corresponds to a calculation for a polycrystalline sample.

\section{Results and discussion}
We have succeeded in synthesizing single-phase Ba$_{1-x}$K$_x$Cd$_2$As$_2$ samples 
over the concentration range $x \leq$ 0.08 (Fig. \ref{XRD}). 
The observed diffraction peaks measured by powder X-ray diffraction can all be indexed 
using the CaAl$_2$Si$_2$-type structure. 
We also tried to synthesize samples for higher K content, but an impurity phase of KCd$_4$As$_3$ 
appears for $x \geq$ 0.12, demonstrating the solid solubility limit. 
The lattice constant $c$ evaluated by the Rietveld method varied linearly with the doping level, 
indicating that K atoms are successfully doped up to $x$ = 0.08 (Fig. \ref{lattice const}). 
The increase with doping is attributed to the larger ionic size of K compared to Ba and 
to the decrease of valency difference between the (Ba,K) ions and the CdAs layers. 
The lattice constant $a$ tends to decrease with doping, indicating that holes are doped into the CdAs covalent bonds. 

A non-doped BaCd$_2$As$_2$ sample exhibits semiconducting behavior 
with the relatively high electrical resistivity of 13.4 $\Omega$m around room temperature, which decreases with heating [Fig. \ref{R_S}(a)]. 
The high resistivity ensures the high purity with low amount of lattice defect of the sample. 
The electrical resistivity decreases rapidly with doping, exhibiting metallic-like behavior at $x$ = 0.08, 
where $\rho$ = 8.5 $\mu$$\Omega$m around room temperature [Fig. \ref{R_S}(b)]. 
We fitted the electrical resistivity for $x$ = 0.00 using the following function to evaluate the band-gap energy ($E\rm_g$) : 
$\rho = \rho_0exp(E\rm_g/2\it k\rm_B\it T)$, 
where $\rho_0$ is a constant and $k\rm_B$ is the Boltzmann constant. 
The resulting value is $E\rm_g$ = 0.40 eV [inset of Fig. \ref{R_S}(b)]. 
A positive value of the Seebeck coefficient indicates a p-type semiconductor [Fig. \ref{R_S}(c)]. 
Non-doped BaCd$_2$As$_2$ exhibits the high Seebeck coefficient $S$ = 527 $\mu$V/K around room temperature and 
reaches a maximum value of 723 $\mu$V/K at 515 K. 
The Seebeck coefficient decreases rapidly with doping to 73 $\mu$V/K around room temperature for $x$ = 0.08. 
The temperature dependence shows a linear increase with heating 
over the whole temperature range for $x$ $\geq$ 0.02, typical of heavily doped semiconductors. 

\begin{figure}
\includegraphics[width=\columnwidth]{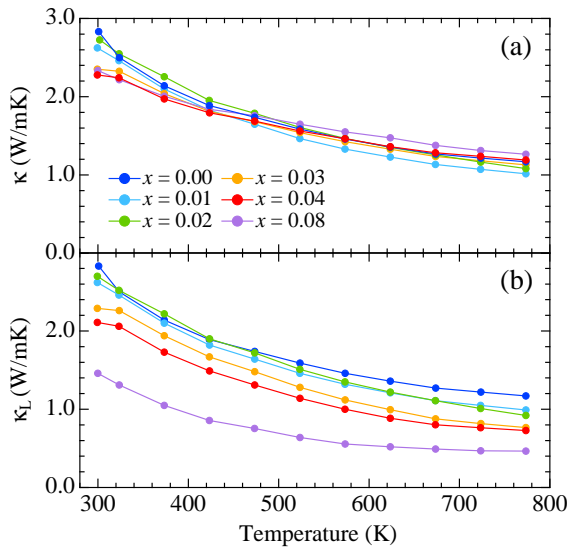}
\caption{\label{Kappa} Temperature dependences of (a) the total thermal conductivity and (b) the lattice thermal conductivity for Ba$_{1-x}$K$_x$Cd$_2$As$_2$.}
\end{figure}

Figure \ref{PF} shows the temperature dependence of the power factor ($S^2/\rho$). 
Around room temperature, the power factor increases with doping, reaching the value 0.63 mW/mK$^2$ for $x$ = 0.08. 
The enhancement is due to the rapid decrease of the electrical resistivity with doping around room temperature. 
The power factor for $x$ = 0.08, however, is lower than that for $x$ = 0.04 at high temperatures, 
due to a smaller decrease of the resistivity with doping. 
The highest value is obtained for $x$ = 0.04, where it increases rapidly with heating up to $T$ $\sim$ 600 K, 
after which it remains almost constant, reaching a maximum value of 1.28 mW/mK$^2$ at 762 K. 

The total thermal conductivity is almost independent of doping [Fig. \ref{Kappa}(a)]. 
The value is $\kappa$ = 2.83 W/mK around room temperature for $x$ = 0.00, 
which decreases with heating to $\kappa$ = 1.17 W/mK at 773 K. 
The lattice thermal conductivity was determined by subtracting the electronic thermal conductivity ($\kappa\rm_e$) from $\kappa$ [Fig. \ref{Kappa}(b)]. 
The Wiedemann-Franz law was applied to calculate $\kappa\rm_e$ 
using $\kappa\rm_e = LT/\rho$, where $L$ = 2.44$\times$10$^{-8}$ W$\Omega$/K$^2$ is the Lorenz number. 
More highly doped samples exhibit larger values of $\kappa\rm_e$ due to their lower electrical resistivity, 
resulting in lower values of $\kappa\rm_L$. 
The lowest value is achieved for $x$ = 0.08: $\kappa\rm_L$ = 1.46 W/mK around room temperature 
and 0.46 W/mK at 773 K. 
The lower values of $\kappa\rm_L$ for the more highly doped samples suggest that the randomness induced by K doping 
effectively scatters the thermal flow. 
The decreases of $\kappa\rm_L$ with heating could be due to an increase in Umklapp scattering of the thermal-transporting acoustic phonons.
The suppression of $\kappa\rm_L$ due to randomness induced by substitution of ionic atoms was also observed in 
Sb-based 122 Zintl compounds, where Ca or Eu atoms are substituted by Yb atoms. \cite{Gascoin_(CaYb)Zn2Sb2,Shuai_(CaEuYb)Zn2Sb2,Zhang_(YbEu)Cd2Sb2,Cao_(CaYb)Cd2Sb2}. 
For example, $\kappa\rm_L$ is about 40\% suppressed by 25\% Yb doping at room temperature in (Yb,Ca)Zn$_2$Sb$_2$. \cite{Shuai_(CaEuYb)Zn2Sb2}
The suppression is, however, milder than that of the present samples. 
This can come from the difference in valency of doped ions.
In contrast that Ca or Eu divalent ions are substituted by the divalent Yb ions, divalent Ba ions are substituted by monovalent K ions in the present samples, 
which can vary the interatomic force constants around ions.
Thus, the effect of randomness can be larger in the present samples.

\begin{figure}
\includegraphics[width=\columnwidth]{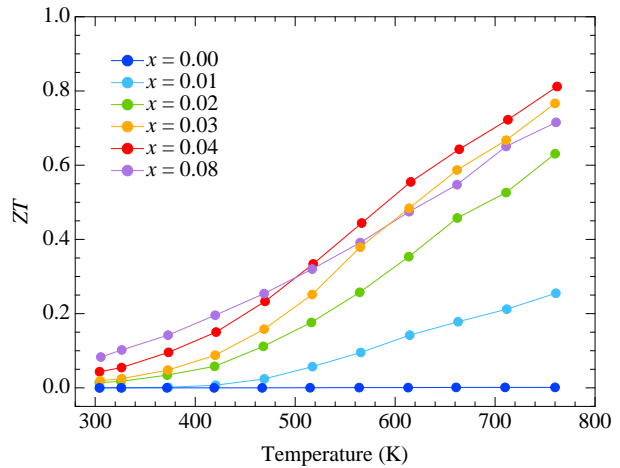}
\caption{\label{ZT} Temperature dependences of $ZT$ for Ba$_{1-x}$K$_x$Cd$_2$As$_2$.}
\end{figure}

Figure \ref{ZT} shows the temperature dependences of $ZT$. 
The $ZT$ values for $x$ = 0.00 are quite low because of the high electrical resistivity. 
They increase with doping up to $x$ = 0.08 around room temperature. 
The doping dependences are similar to those for the power factor because the total thermal conductivity is independent of doping. 
The highest $ZT$ value is obtained with $x$ = 0.04 sample, where it increases with heating 
to a maximum value of 0.81 at 762 K. 

\begin{figure}
\includegraphics[width=\columnwidth]{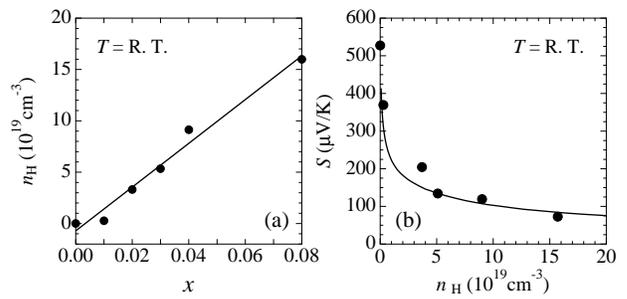}
\caption{\label{S_n} (a) Hall carrier concentration vs. K content. The solid line is a linear fit. 
(b) Seebeck coefficient vs. Hall carrier concentration around room temperature. 
The solid line is a calculated curve based on Boltzmann transport theory using the band structure obtained from the WIEN2k package.}
\end{figure}

The hall carrier concentration ($n\rm_H$) at room temperature increases linearly with $x$, 
indicating the successive hole doping in the CdAs covalent bonding layers results from the substitution of K for Ba [Fig. \ref{S_n}(a)]. 
The value of $n\rm_H$ for $x$ = 0.00 is quite low, which is 3.15$\times10^{16}$ cm$^{-3}$, 
demonstrating the low impurity and deficiency levels in the prepared samples. 
The value of $n\rm_H$ reaches 1.60$\times10^{20}$ cm$^{-3}$ for $x$ = 0.08, 
covering the range of carrier concentrations suitable for thermoelectric properties. 
The Hall mobility, given by $\mu\rm_H = 1/(\it n\rm_H\it e\rho)$, 
exhibits the relatively low values of 0.15 and 46 cm$^2$/Vs at room temperature for $x$ = 0.00 and 0.08, respectively. 
The increase of $\mu\rm_H$ with doping can be attributed to the variation of conduction mechanism from hopping to 
metallic like conduction.

\begin{figure}[h]
\centering
  \includegraphics[height=8cm]{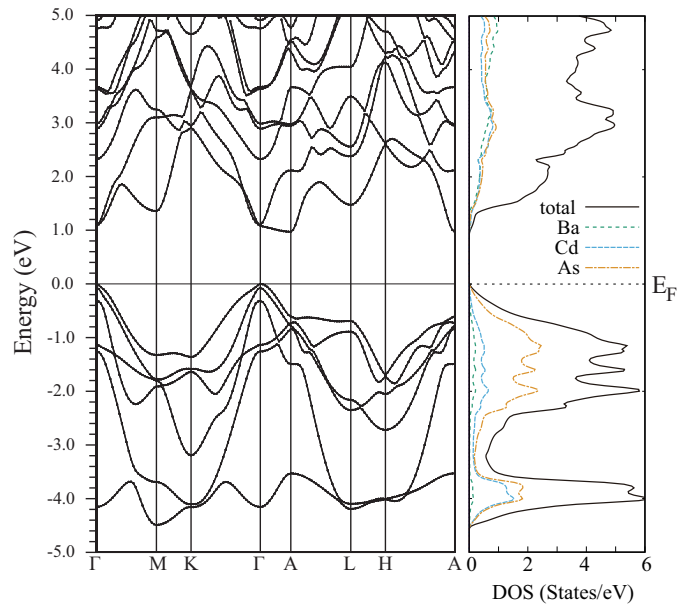}
  \caption{The band structure and the density of states for BaCd$_2$As$_2$.}
  \label{band}
\end{figure}

The band structure and the density of states for BaCd$_2$As$_2$ obtained by using the experimentally determined crystal structure at $x$ = 0.00 are shown in Fig. \ref{band}. 
The valence bands are mainly constructed from the As $p$ orbitals, and the highest and second highest valence bands at the $\Gamma$ point near the Fermi level originate from the As $p_x$ and $p_y$ orbitals.
The two bands constructed from the As $p_x$ and $p_y$ orbitals are split due to the spin orbit coupling.
The energy difference between the two bands is about 0.07 eV, which is similar to those of other As- and Sb-based 122 Zintl phase compounds \cite{Zhang_122calc,Singh,Sun}. 
The band structure originating from the As $p_z$ orbital lies below the Fermi level and the energy difference $\Delta_p$ between the top of the valence band and $p_z$ bands is about 0.36 eV.
This is relatively large compared to those of other 122 Zintl compounds discussed in reference \cite{Zhang_122calc}.

The calculated Seebeck coefficient as a function of the carrier concentration is in accordance with the experimental result around room temperature [Fig.\ref{S_n}(b)], which suggests that the calculated electronic structure basically reproduces that of the actual material.
The bands lying within an energy range of several $k\rm_B\it T$ from the Fermi level contribute to transport coefficients such as the electrical conductivity and the Seebeck coefficient.
Because $\Delta_p$ is about $14k\rm_B\it T$ at 300 K, we conclude that the $p_{x,y}$ bands provide the main contributions to the observed transport properties, while the $p_z$ band makes only a small contribution.

As discussed in reference \cite{Zhang_122calc}, smaller values of $\Delta_p$, namely enhanced band degeneracy, are preferable for producing larger maximum power factors in this series of materials. 
The maximum of the experimentally determined power factor for the present materials ($1.28$ mW/mK$^2$ at 762 K) is relatively large; 
it is larger than that of other 122 Zintl compounds having similar values of $\Delta_p$. 
To investigate the origin of the large power factor for BaCd$_2$As$_2$, we have evaluated the effective masses in the $x$, $y$, and $z$ axis directions, $m_{xx}$, $m_{yy}$, and $m_{zz}$, respectively.
The anisotropy of the effective masses reflects the dimensionality of the band structure, which is known to strongly affect the maximum power factor, as discussed in, e.g., references \cite{Hicks,Parker,Gibbs,Sun}. 
The calculated effective masses at the top of the valence band are $m_{xx} =0.49m_e$, $m_{yy} =0.83m_e$, and $m_{zz} =0.65m_e$, where $m_e$ is the electron rest mass.
For the purpose of comparison, we have also calculated the effective mass and the power factor for CaZn$_2$As$_2$ using a theoretically optimized lattice structure with the PBEsol exchange-correlation functional\cite{Perdew2008}, where the energy splitting between the $p_x$ and $p_y$ bands is almost the same (0.06eV) and the $p_{z}$ orbital similarly makes a small contribution to the Seebeck effect ($\Delta_p$ = 0.31 eV). 
It is found that the effective masses for CaZn$_2$As$_2$ are $m_{xx}=0.42m_e$, $m_{yy}=0.58m_e$, and $m_{zz}=0.60m_e$, and the calculated power factor for BaCd$_2$As$_2$  at around $n\rm_H$ = 1 $\times 10^{20}$ cm$^{-3}$ is about 1.4 times larger than that of CaZn$_2$As$_2$ at 300 K assuming the same relaxation times for these materials.
The difference in the effective mass of the second-highest valence band is negligible between the two materials in comparison with that of the highest valence band.
Considering the large difference in $m_{yy}$ between the two materials, we conclude that the relatively large power factor of BaCd$_2$As$_2$ originates from the anisotropy of the band structure, in addition to the approximate two-fold degeneracy ($p_x$ and $p_y$) of the bands.
As the present material has a relatively large value of $\Delta_p$, we anticipate that there is more room left for further enhancement of the power factor by enhancing the band degeneracy around the Fermi energy, which may be accomplished by element substitution.

\section{Conclusions}
Single-phase samples of Ba$_{1-x}$K$_x$Cd$_2$As$_2$ (0.00 $\leq x \leq$ 0.08) were successfully synthesized using solid-state reactions. 
The band-gap energy for $x$ = 0.00 was determined to be 0.40 eV from electrical resistivity measurements. 
The maximum power factor is 1.28 mW/mK$^2$ at 762 K for $x$ = 0.04. 
This relatively large power factor can be due to the anisotropic band structure and the two-fold degeneracy at the valence-band maximum. 
The total thermal conductivity is almost independent of doping, taking the value 1.17 W/mK at 773 K for $x$ = 0.00. 
The highest value of $ZT$ obtained is 0.81 at 762 K for $x$ = 0.04.

\section*{Conflicts of interest}
There are no conflicts to declare.

\section*{Acknowledgement}
We would like to thank K. Suekuni for valuable discussions and M. Ohta for supporting Hall coefficient measurements.
This work was supported by CREST (No. JPMJCR16Q6) from the Japan Science and Technology Agency and 
by the New Energy and Industrial Technology Development Organization through the Thermal Management Materials and Technology Research Association.

\end{document}